\documentclass[twocolumn,amsmath,ammsymb]{revtex4-1}
\usepackage{epsfig}
\usepackage{amsmath}
\usepackage{amssymb}
\usepackage{graphicx}
\usepackage{dcolumn}
\usepackage{soul}
\usepackage{bm}
\usepackage[usenames]{color}

\begin{document}

\preprint{}

\title{Interaction-induced phase transitions of type-II Weyl semimetals}
\author{Yi-Xiang Wang$^1$, Fuxiang Li$^2$, and Baoan Bian$^1$}
\affiliation{$^1$School of Science, Jiangnan University, Wuxi 214122, China.}
\affiliation{$^2$Center for Nonlinear Studies and Theoretical Division, Los Alamos National Laboratory, Los Alamos, NM 87545 USA.}

\date{\today}

\begin{abstract}
The study of Weyl semimetal (WSM) lies at the forefront of the nontrivial topological phenomena in condensed matter physics.  In this work, we study the effect of onsite repulsive Hubbard interaction on the WSM system with a nonzero tilt at half-filling.  Within the Hartree-Fock mean-field (MF) approximation, we treat the Hubbard interaction self-consistently and find that the Fock exchange field vanishes while the Hartree field can renormalize the topological mass, the tilt and the Fermi velocity of the Weyl cones.  When the renormalized tilt is larger than the renormalized Fermi velocity, the Hubbard interaction will induce the quantum phase transition from type-I WSM to type-II WSM.  We then provide the interaction-induced phase diagrams of WSM in different parametric spaces, in which the antiferromagnetic order at strong interaction is also considered.  In addition, we analyze another model hosting two pairs of Weyl nodes and similar results are obtained.  The implications of these results are discussed.
\end{abstract}


\maketitle

\section{\label{sec:level1} Introduction}

Since the discovery of topological insulator \cite{M.Z.Hasan,X.L.Qi}, the conventional band theory of solids was dramatically revolutionized.  The topologically nontrivial quadratic Hamiltonians have been extended to the three-dimensional (3D) Weyl semimetals (WSMs).  One of the most important features that the WSMs bring to this area is that they are gapless states of matter, which are topologically nontrivial and whose realizations are of significant importance, just as the gapped topological insulators.  The theoretical proposals for the Weyl nodes in the band structure of solid state materials require breaking either inversion symmetry or time-reversal symmetry (TRS), resulting in the separation of a pair of Dirac nodes into Weyl nodes with opposite chiralities \cite{D.Hsieh,A.A.Burkov,X.Wan,G.Xu,G.Volovik}.  The ideal WSM has a conical spectrum and a point-like Fermi surface at the Weyl node.  When the strain or chemical doping is present, the energy dispersion in the momentum space at a Weyl node could generally be tilted along a certain direction.  If the tilt is small that the Fermi surface remains point-like, the system is classified as type-I WSM (WSM1).  When the tilt becomes large enough, the Fermi surface may no longer remain as point-like, but instead consists of electron and hole pockets.  In this case, the system is called as type-II WSM (WSM2) \cite{A.A.Soluyanov}.  Besides the inversion symmetry or TRS, WSM2 additionally breaks the Lorentz invariance.

Initially, WTe$_2$ was predicted by \textit{ab} initio calculations to be a possible candidate for the experimental realization of WSM2 \cite{A.A.Soluyanov}.  Later, MoP$_2$ and WP$_2$ were predicted to host four pairs of type-II Weyl points and own long topological Fermi arcs, which make them readily accessible in angle-resolved photoemission spectroscopy (ARPES) \cite{G.Autes}.  Meanwhile the transport and thermodynamical properties of WSM2 are evidently different from WSM1 and have attracted many interests in theory, such as the field-selective anomaly in magnetotransport \cite{Z.M.Yu,M.Udagawa,S.Tchoumakov}, the intrinsic anomalous Hall effect \cite{A.A.Zyuzin} and the tilt-dependent optical conductivity \cite{J.P.Carbotte}.  These can be attributed to the overtilted Weyl cones and the finite density of states at the Fermi level of WSM2.  In a recent work, Park \emph{et.al} considered the possibility of disorder-induced WSM1-WSM2 transition in the framework of Born approximation, which provides a possible route to realize the WSM2 phase \cite{M.J.Park}.  There are also several experimental progresses, reporting the ARPES and scanning tunnelling microscopy (STM) evidences of WSM2 in MoTe$_2$ \cite{L.Huang,K.Deng,J.Jiang,N.Xu}, Mo$_x$W$_{1-x}$Te$_2$ \cite{I.Beloposki} and LaAlGe \cite{S.Y.Xu}.

It is well known that when the 2D topological states are combined with the Hubbard interaction, the interplay between the correlation and band topology can drive the system towards different electronic orders \cite{W.Zheng,D.Prychynenko,A.M.Cook,V.S.Arun,J.He1,J.He2,T.I.Vanhala,Y.X.Wang}.  The correlation effects in 3D WSM are worth exploring as well \cite{J.Liu,B.Roy}.  In nodal loop semimetals, it has been found that the Hubbard interaction can induce the surface ferromagnetic phase through the continuous quantum phase transition, while the bulk remains robust against local interaction and nonordered \cite{J.Liu}.  In another work of line-node semimetal, either the antiferromagnetic order or charge density wave dominates the system, depending on the relative strength of onsite and nearest-neighbor repulsions \cite{B.Roy}.  Motivated by these progresses, in this work we will study the effect of repulsive Hubbard interaction in inducing the quantum phase transitions in WSM with a nonzero tilt.

Within the Hartree-Fock mean-field (MF) approximation, we treat the on-site Hubbard interaction on 3D Weyl fermion self-consistently.  The main results obtained are as follows: (a) At the MF level, the Fock exchange field vanishes while the Hartree field can renormalize the topological mass.  We find the local magnetization plays a key role in determining the topological phase transitions and its magnitude is strengthened by the Hubbard interaction.  We also perform a detailed analysis of how the local magnetization is related to the effective magnetic field and the tilt of the cone.  (b) When the renormalized tilt is larger than the renormalized Fermi velocity, the interaction-induced quantum phase transition from WSM1 to WSM2 will occur.  Based on these results, the interaction-induced phase diagrams are obtained in different parametric spaces.  The effect of thermal fluctuations is also analyzed.  We suggest that the Hubbard interaction can provide an effective route in driving the phase transitions to WSM2.  (c) The antiferromagnetic (AFM) order is studied within an enlarged unit cell and it can appear when the Hubbard interaction is strong enough.  (d) We analyze the model hosting two pairs of Weyl nodes and similar results are obtained.  We hope our work can provide some insights into the understanding about the competition between the correlation and topology in 3D WSM.  The interaction-driven WSM2 phase may be of particular interests for semiconductor technology in the future.

\section{\label{sec:level1} Noninteracting Model}

\begin{figure}
\includegraphics[width=8.8cm]{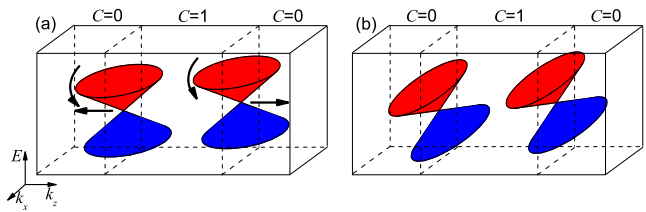}
\caption{(Color online) Schematic plot of the tilting Weyl cones in the $k_x-k_z$ plane of WSM1 with $\gamma_z=0.4$ in (a) and WSM2 with $\gamma_z=1.2$ in (b).  The Chern numbers are shown in different layers.  As shown in (a), when the Hubbard interaction-induced the renormalized topological mass $m_1'$ increases, the Weyl nodes will move to the edge of the BZ and the Weyl cones will get more tilted.  Note the energies of Weyl nodes are unequal, leading to the existence of electron and hole Fermi surfaces at half-filling. }
\end{figure}

We start from the spinful Hamiltonian $H=H_0+H_t$ describing a pair of Weyl fermions (the lattice constant is set as $a=1$) \cite{M.J.Park,H.Shapourian}:
\begin{eqnarray}
H_0&=&t(\text{sin}k_x\sigma_x+\text{sin}k_y\sigma_y)+(m_1+t\text{cos}k_z)\sigma_z
+m_0(2
\nonumber\\
&&-\text{cos}k_x-\text{cos}k_y)\sigma_z-\mu\sigma_0,
\nonumber\\
H_t&=&a_t\text{sin}k_z\sigma_0.
\end{eqnarray}
Here $\bm\sigma$'s are the Pauli matrices denoting the spin-$1/2$ degree of freedom. $t$ and $m_1$ are the hopping integral and topological mass, respectively.  When $|m_1|<t$, the Weyl cones are located at ${\bf K}_\pm=(0,0,\pm Q)$ in the 3D Brillouin zone (BZ), where $Q=\text{arccos}(-\frac{m_1}{t})>0$.  $\mu$ is the chemical potential of the system.  The term of Wilson mass $m_0$ assures the stability of the Weyl cones.  $H_0$ preserves the inversion symmetry ${\mathcal I}^{-1}H_0({\bf k}){\mathcal I}=H_0(-{\bf k})$ with the inversion operator ${\mathcal I}=\sigma_z$ but breaks the TRS with the time-reversal operator defined as ${\mathcal T}={\mathcal K}$ and ${\mathcal K}$ the complex conjugation operator \cite{T.M.McCormick}.  $H_t$ specifies the tilt in the $z-$axis direction.  Such a tilting term that is odd in momentum and breaks the inversion symmetry was analyzed and discussed in the context of WSM2 in previous works \cite{A.A.Soluyanov,K.Deng,M.Udagawa,Y.Wang}.  A similar two-band model was shown to emerge from a topological insulator-normal insulator (TI-NI) heterostructure \cite{A.A.Burkov}, and the tilting term can be generated by including the spin-orbit coupling (SOC) between the TI-NI interfaces \cite{A.A.Zyuzin2}.  In the following, we will use $t$ as the unit of energy.

The low-energy Hamiltonian can be obtained by expanding $H$ around the Weyl nodes ${\bf K}_\pm$ as ($\hbar=1$):
\begin{eqnarray}
H_\pm({\bf q})=v(q_x\sigma_x+q_y\sigma_y)\mp v_z q_z\sigma_z+(\gamma_z q_z\pm c_0)\sigma_0,
\end{eqnarray}
with the momentum ${\bf q}={\bf k}-{\bf K}_\pm$ being the deviation from the Weyl nodes.  The Fermi velocities are given as $v=t$ and $v_z=t\text{sin}Q$, so the Weyl cones are generally not isotropic.  The tilting factor is given by $\gamma_z=a_t\text{cos}Q$ and the constant term $c_0=a_t\text{sin}Q$.  When the tilting factor becomes larger than the Fermi velocity in the same direction, $\gamma_z>|v_z|$, the system enters the WSM2 phase.  In Fig. 1, the schematic plots of the tilting Weyl cones in $k_x-k_z$ plane are shown of the WSM1 in (a) and WSM2 in (b).

The Fermi arc that links the projection of the bulk Weyl points with opposite  chiralities in the surface BZ is one of the most prominent features of the WSM \cite{X.Wan}.  Consider a slab of WSM that is infinite in the $x-$ and $z-$directions while semi-infinite in the $y-$direction, filling the $y>0$ half-plane.  The energy eigenvalue problem in the real space is ${\mathcal H}_\pm(q_x,-i\partial_y,q_z)\Psi_{s\pm}({\bf r})=E_{s\pm}(q_x,q_z)\Psi_{s\pm}({\bf r})$, where the Hamiltonian around the Weyl node ${\bf K}_\pm$ is:
\begin{eqnarray}
&&{\mathcal H}_\pm(q_x,-i\partial_y,q_z)
\nonumber\\
&&=v_x q_x\sigma_x-iv_y\partial_y\sigma_y
\mp v_z q_z\sigma_z+(\gamma_z q_z\pm c_0)\sigma_0+M(y)\sigma_z.
\nonumber\\
\end{eqnarray}
Here to model the boundary, we take $M(y)=M$ for $y<0$ and $M(y)=0$ for $y>0$ \cite{P.Goswami}.  Taking the limit of $M\rightarrow\infty$ models the interface with vacuum or a large-gap trivial insulator.  It can be shown that only the state corresponding to eigenvalue $+1$ of the matrix $\sigma_x$ can lead to the normalizable solution.  So the eigenenergy is
\begin{eqnarray}
E_{s\pm}(q_x,q_z)=v_x q_x+(\gamma_z q_z\pm c_0 ),
\end{eqnarray}
and the corresponding wavefunction of
\begin{eqnarray}
\Psi_{s\pm}({\bf r})=\sqrt{\frac{q_z}{2}}e^{iq_xx \mp iq_zz}e^{-\frac{v_z}{v_y}q_z y} \begin{pmatrix}1\\ 1\end{pmatrix},
\end{eqnarray}
with $q_z>0$.  Eq. (4) tells us that the surface states at different Weyl nodes have certain energy difference if $a_t\neq0$.  The linear characteristic of the surface states is in good accordance with the ARPES measurement in MoTe$_2$ \cite{L.Huang}.

In Hamiltonian $H_\pm$, the tilting factor $\gamma_z$ and the Fermi velocity $v_z$ in $z-$direction are strongly dependent on the topological mass $m_1$ and tilting parameter $a_t$.  Thus the change of $m_1$ and $a_t$ may drive the system enter different phases.  The phase diagram of noninteracting WSM in the parametric space of $m_1$ and $a_t$ is shown in Fig. 2.  One can clearly see that there exist two kinds of phase transition \cite{M.J.Park}: the metal-insulator transitions and the WSM1-WSM2 continuous transitions, where the phase boundaries are shown with the dashed and dotted lines, respectively.

\begin{figure}
\includegraphics[width=9.2cm]{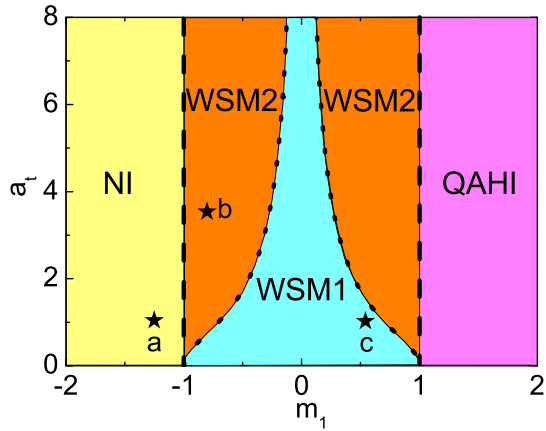}
\caption{(Color online) Phase diagram of the WSM in the noninteracting case with $m_0=2$, where the different phases are shown with different colors.  The phase boundary of solid lines characterize the metal-insulator topological phase transition while the dotted lines describes the WSM1-WSM2 continuous transitions.  The stars a-c are the initial phases for interaction-induced phase transitions, as denoted by the arrows in Fig. 3. }
\end{figure}

These different phases can be characterized by the nontrivial Hall conductance, which is obtained from the famous TKKN formula \cite{D.J.Thouless}.  When $|m_1|<t$, there exists a pair of gapless Weyl nodes.  To calculate the Hall conductance, the 3D system can be considered as the stacking of the 2D slices at each momentum $k_z$.  Each slice describes the 2D gapped Dirac fermions in $k_x-k_y$ plane with mass $m_\pm(k_z)=\mp v_z(k_z\mp Q)$ around the Weyl nodes ${\bf K}_\pm$ and the mass vanishes at ${\bf K}_\pm$.  It should be noted that the Dirac fermions around ${\bf K}_\pm$ own the same chiralities.  The total Hall conductance $\sigma_H$ of the 3D system is a summation over each slice and is given as \cite{A.A.Burkov}:
\begin{eqnarray}
\sigma_H=\frac{e^2}{2\pi h}\sum_{k_z\in\text{BZ}}[\text{sgn}(k_z+Q)-\text{sgn}(k_z-Q)].
\end{eqnarray}
As shown in Fig. 1, only topological nontrivial layers with Chern number $C=1$ in the middle region $(-Q<k_z<Q)$ contribute to $\sigma_H$, while the topological trivial layers with $C=0$ in the left $(k_z<-Q)$ or right $(k_z>Q)$ region make no contribution to $\sigma_H$.  The corresponding Hall conductance of the system is given by $\sigma_H=\frac{Qe^2}{\pi h}$, i.e., proportional to the separation between the two Weyl nodes.  When $|m_1|>t$, the two Weyl nodes meet and annihilate so that the system becomes gapped and thus enters the insulator phase.  Especially for the case of $m_1>t$, the Weyl nodes annihilate at $(0,0,\pm\pi)$, the boundary of the BZ, leading to $\sigma_H=\frac{e^2}{h}$ and the system enters the quantum anomalous Hall insulator (QAHI) phase.  For the case of $m_1<-t$, the Weyl nodes annihilate at $(0,0,0)$, the center of the BZ, leading to $\sigma_H=0$ and the system enters the normal insulator (NI) phase.

The transition from WSM1 to WSM2 at nonzero $m_1$ happens when the tilting factor $\gamma_z$ increases to be larger than the Fermi velocity in the $z-$direction, $\gamma_z>|v_z|$, i.e.,
\begin{eqnarray}
a_t\Big|\frac{m_1}{t}\Big|>t\sqrt{1-\Big(\frac{m_1}{t}\Big)^2},
\end{eqnarray}
which shows the phase boundary is nonlinear.  As $m_1$ approaches to zero, the phase boundary of WSM1-WSM2 transitions extends to infinity.

\section{mean-field theory}

We consider the half-filling case, i.e., there is only one electron on each site.  This can be achieved by modulating the chemical potential in the system.  As the energies of Weyl nodes are unequal, the electron and hole Fermi surfaces can coexist.  Then the long-range Coulomb interaction are expected to be effectively screened by the finite density of states at half-filling and can instead be described by the on-site Hubbard interaction \cite{G.Y.Cho,C.Chan}:
\begin{eqnarray}
H_U=U\sum_ln_{l\uparrow}n_{l\downarrow},
\end{eqnarray}
here $U>0$ is the repulsive interaction strength and $n_{l\alpha}=c_{l\alpha}^+c_{l\alpha}$ denotes the electron number at site $l$ with spin $\alpha$.  When the interaction is strong and much larger than the energy scale of the system, $U\gg t$, it is evident that the ground state of the system is a charge-localized Mott insulator \cite{V.S.Arun}.  While for intermediate interaction strength, $U\sim t$, the correlation effect between electrons will compete with the topology of the bands.

To decouple the local Hubbard interaction, we apply the Hartree-Fock MF approximation, with all possible channels included.  The MF theory has been successfully applied in several fields of strongly correlated electrons \cite{J.He1,J.He2,A.M.Cook,W.Zheng,T.I.Vanhala,V.S.Arun,D.Prychynenko,Y.X.Wang,J.Liu,B.Roy}.  Previously, we applied the MF theory in two spatial dimensions to investigate the topological phase transitions in the arbitrary Chern number insulator \cite{Y.X.Wang}.  Here we further extend the MF theory and consider the 3D Weyl system.

We define the MF parameters of local charge density $\rho_l=\sum_\alpha\langle n_{l\alpha}\rangle$ and local magnetization ${\cal M}_l=\sum_{\alpha\beta}\langle c_{l\alpha}^+{\bm\sigma}_{\alpha\beta}c_{l\beta}\rangle$ \cite{V.S.Arun}.  It should be noted that ${\cal M}_{lz}$ is not a symmetry-breaking order parameter, but only leads to the shift of the quantum critical points at which the energy bands become gapless.  ${\cal M}_{l-}={\cal M}_{lx}-i{\cal M}_{ly}$ can act as a symmetry-breaking order parameter, whose nonvanishing value will lead to a spontaneous nematic order, suggesting that the lattice rotational symmetry around $z-$direction is broken \cite{A.M.Cook}.  With the help of these MF parameters, the Hubbard term can be decoupled as:
\begin{eqnarray}
H_U^d&=&\frac{U}{2}\sum_l\rho_l\sum_\sigma c_{l\sigma}^+c_{l\sigma}-\frac{U}{2}
\sum_l{\cal M}_l\cdot{\bm\sigma}_l,
\end{eqnarray}
in which ${\bm\sigma}_l=(\sigma_{lx},\sigma_{ly},\sigma_{lz})$ are the Pauli matrices representing an electron's spin at site $l$.  We have dropped the constant terms in Eq. (9).

\section{\label{sec:level1} Main Results}

\subsection{Renormalized topological mass}

First we study the renormalization of topological mass and ignore any kind of many-body instabilities.  Under this assumption, the bulk system possesses the translational symmetry and the MF parameters should be spatially uniform.  So in the following we use $\rho$ and $\cal M$ to represent the local $\rho_l$ and ${\cal M}_l$.

In the framework of MF theory, the Hubbard interaction will modify the original noninteracting Hamiltonian to the MF Hamiltonian $H_{\text{mf}}$, which in momentum space takes the following form:
\begin{eqnarray}
H_{\text{mf}}({\bf k})
&=&\begin{pmatrix}
A_{\bf k}-\frac{U}{2} {\cal M}_z & B_{\bf k}-\frac{U}{2}{\cal M}_-\\
B_{\bf k}^*-\frac{U}{2}{\cal M}_+&
-A_{\bf k}+\frac{U}{2} {\cal M}_z
\end{pmatrix}
+(a_t\text{sin}k_z+\frac{U}{2}\rho)\sigma_0.
\nonumber\\
\end{eqnarray}
Here the variables in the matrix are $A_{\bf k}=m_1-t\text{cos}k_z+m_0(2-\text{cos}k_x-\text{cos}k_y)$, $B_{\bf k}=t(\text{sin}k_x-i\text{sin}k_y)$.  The eigenenergies are given as $\varepsilon_\pm({\bf k})=\pm D_{\bf k}+a_t\text{sin}k_z+\frac{U}{2}\rho$, where $D_{\bf k}=\sqrt{(A_{\bf k}-\frac{U}{2}{\cal M}_z)^2+|B_{\bf k}-\frac{U}{2}{\cal M}_-|^2}$.  It is clear that the term of $\frac{U}{2}\rho$ in the eigenenergies shifts the energy level by $\frac{U}{2}\rho$,  while another term of $a_t$sin$k_z$ does not.

Using the eigenengies and eigenstates of $H_{\text{mf}}({\bf k})$, the self-consistent equations for $\rho$ and $\cal M$ are
\begin{eqnarray}
&&\rho=\frac{1}{N}\sum_{\bf k}\Big\{f[\varepsilon_+({\bf k})]+f[\varepsilon_-({\bf k})]\Big\},
\\
&&{\cal M}_-=\frac{1}{N}\sum_{\bf k}\frac{B_{\bf k}-\frac{U}{2}{\cal M}_-}{D_{\bf k}} \Big\{f[\varepsilon_+({\bf k})]-f[\varepsilon_-({\bf k})]\Big\},
\\
&&{\cal M}_z=\frac{1}{N}\sum_{\bf k}\frac{A_{\bf k}-\frac{U}{2}{\cal M}_z}{D_{\bf k}} \Big\{f[\varepsilon_+({\bf k})]-f[\varepsilon_-({\bf k})]\Big\},
\end{eqnarray}
here $f(\varepsilon_\alpha)=1/(e^{\beta(\varepsilon_\alpha-\mu)}+1)$ is the Fermi distribution function with the energy $\varepsilon_\alpha({\bf k})$ and inverse temperature $\beta=\frac{1}{k_BT}$.  The chemical potential is set as $\frac{U}{2}\rho$ to keep the system half-filling.  We solve the equations by the self-consistent iterative approach \cite{V.S.Arun,Y.X.Wang,D.Prychynenko}.  The steps are as follows: (a) set initial random values for $\rho$ and $\cal M$; (b) diagonalize $H_{\text{mf}}({\bf k})$ as to solve the energies and eigenvectors; (c) use the obtained energies and eigenvectors to calculate new $\rho$ and $\cal M$.  Repeat these steps until convergence is reached.  The convergence conditions are set to be $|\Delta\rho|<10^{-6}$ and $|\Delta{\cal M}|<10^{-6}$, where $\Delta\rho$ and $\Delta{\cal M}$ are the differences in $\rho$ and $\cal M$ between the subsequent iterations, respectively.  As further checks on the numerical results, we set different initial values for the MF parameters and find the results exhibit good convergence.  In fact in the case of half-filling, the local charge density gives as $\rho=1$.

The calculation shows that in the zero-temperature case, ${\cal M}_-$ vanishes, suggesting that there is no rotational-symmetry breaking and no nematic phase.  This can be explained from Eq. (12) as follows.  At zero temperature, the nonzero contributions must come from the electronic states satisfying $f[\varepsilon_{+}({\bf k})]=0$ and $f[\varepsilon_{-}({\bf k})]=1$, which requires the condition of
\begin{eqnarray}
D_{\bf k}>a_t|\text{sin}k_z|.
\end{eqnarray}
If the tilting parameter $a_t$ is small, this condition is automatically satisfied for all momenta in the BZ.  If the tilting parameter $a_t$ is large, the allowed momentum space is reduced but is still symmetric with $z-$axis: $(k_x,k_y,k_z)\leftrightarrow(-k_x,-k_y,k_z)$.  Then Eq. (12) can be rewritten as:
\begin{eqnarray}
{\cal M}_-=\frac{\frac{1}{N}\sum_{\bf k}\frac{B_{\bf k}}{D_{\bf k}}}
{\frac{U}{2N}\sum_{\bf k}\frac{1}{D_{\bf k}}-1}.
\end{eqnarray}
As the variables have the properties of $B(k_x,k_y)=-B(-k_x,-k_y)$ and $D(k_x,k_y)=D(-k_x,-k_y)$, when summing the allowed momenta over the BZ, the contribution from ${\bf k}_1=(k_x,k_y,k_z)$ and ${\bf k}_2=(-k_x,-k_y,k_z)$ will exactly cancel with each other, leading to the vanishing of ${\cal M}_-$.  This conclusion can be extended to the finite temperature $T>0$ case.  The above analysis leads to the fact that the MF approximation here is equivalent to adding a Hartree field to the topological mass $m_1$, which is renormalized as
\begin{eqnarray}
m_1'=m_1-\frac{U}{2}{\cal M}_z.
\end{eqnarray}

\begin{figure}
\includegraphics[width=9.2cm]{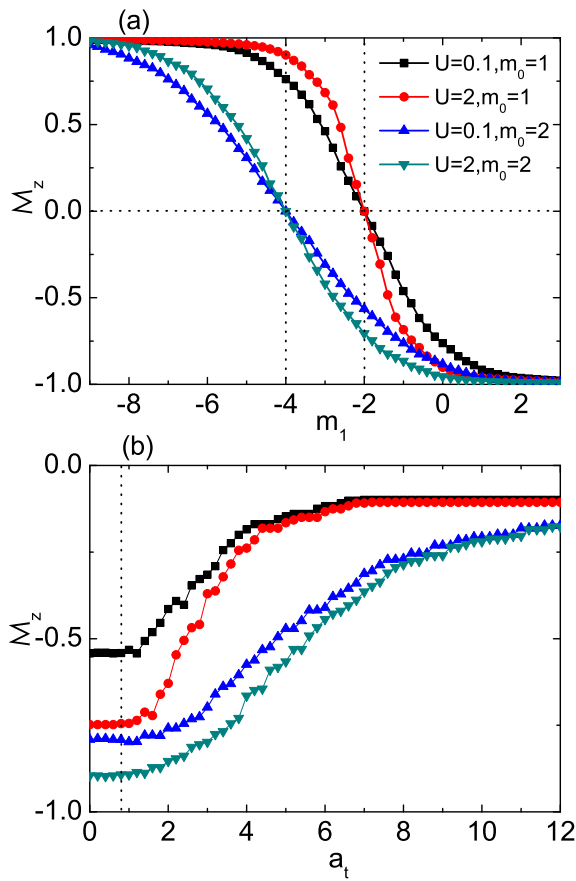}
\caption{(Color online) The magnetization ${\cal M}_z$ vs the topological mass $m_1$ in (a) and the tilting parameter $a_t$ in (b) at different Hubbard interaction strength and Wilson mass $(U,m_0)$.  We have fixed $a_t=1$ in (a) and $m_1=-0.8$ in (b).  The legends are the same in both figures.}
\end{figure}

Finite magnetization ${\cal M}_z$ indicates the existence of ferromagnetic order in the system.  In fact, when on average over the momentum space, the Hamiltonian in Eq. (1) leads to total energy ${\cal E}\sim h_e{\cal M}_z$, in which we define the effective magnetic field $h_e=m_1+2m_0$.  If the the effective magnetic field is negative $h_e<0$, to minimize energy, ${\cal M}_z>0$.  And vice versa.  This leads to another observation that the role Wilson mass term $m_0$ is twofold: it can not only stabilize the Weyl nodes at ${\bf K}_\pm$, but also provide part of the effective magnetic field.  Numerical calculation also verifies this conclusion.  Indeed, in Fig. 3(a) we plot ${\cal M}_z$ vs $m_1$ for several sets of parameters $(U,m_0)$ with fixed $a_t=1$.  For each curve as $m_1$ increases, ${\cal M}_z$ decreases, from the saturation value $+1$ (when $h_e<-4$) to another saturation value $-1$ (when $h_e>4$).  In particular, at $h_e=0$, ${\cal M}_z$ vanishes due to the electrons being equally distributed between the two spin states.  More importantly, it is shown that the existence of ferromagnetic order will get enhanced when the interaction increases.  This is because the corresponding Hartree field strengthens the effective magnetic field \cite{T.I.Vanhala}, leading to larger $|{\cal M}_z|$.

\begin{figure}
\includegraphics[width=9.2cm]{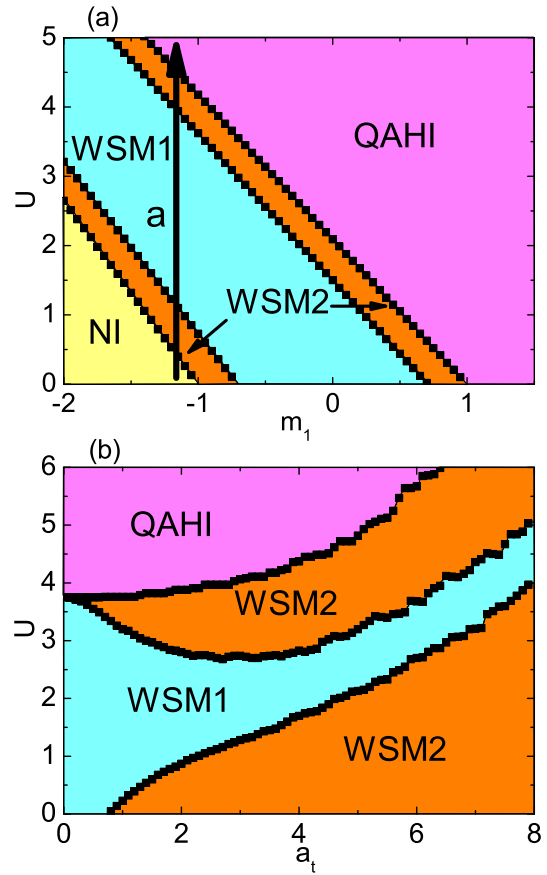}
\caption{(Color online) Interaction-induced phase diagrams of WSM.  (a) is shown in the parametric space of $(U,m_1)$ with $a_t=1$ and (b) is shown in $(U,a_t)$ with $m_1=-0.8$.  The different phases are shown in different colors.  Note the linear phase boundaries in (a) and the nonlinear phase boundaries in (b).  The insets in (a) and (b) show the phase diagram of FM-AFM transitions at large $U$.}
\end{figure}

Furthermore, ${\cal M}_z$ is also dependent on the tilting parameter $a_t$, as shown in the numerical results in Fig. 3(b) with fixed $m_1=-0.8$.  The behavior can be explained as follows.  If $a_t$ is small, the allowed states are unaffected, just as the non-tilting case.  So ${\cal M}_z$ keeps almost unchanged and the boundary is shown by the dotted line in Fig. 3(b).  If $a_t$ is large, the allowed states are reduced, resulting in the decreasing of $|{\cal M}_z|$.  When the tilting parameter increases to be too large, $a_t\gg t$, the tilting term of $a_t\text{sin}k_z$ will dominate in the MF Hamiltonian.  In this case, only the electronic states at $k_z=0,\pm\pi$ plane in the BZ will make contribution to ${\cal M}_z$, at which the tilting term vanishes.  Thus ${\cal M}_z$ gradually reaches its saturation value when $a_t$ becomes large, as shown in Fig. 3(b).

The topological mass $m_1$ controls the metal-insulator topological transitions and can be regulated by external means in experiment, for example, in TI-NI heterostructure \cite{A.A.Zyuzin,A.A.Burkov}, by tuning the thickness of each layer or the concentration of magnetic impurities. For the study of topological phases transitions, the meaningful range of $m_1$ is of the same order of magnitude with $t$. Therefore, the argument of the validity of mean field theory is still applicable in the presence of $m_1$ and, for similar reasons, of tilting parameter $a_t$.

So far we have demonstrated that in the MF theory, the renormalized topological mass due to the Hubbard interaction shows complex behavior: it can be increased when the effective magnetic field is negative or decreased when the effective magnetic field is positive.  This is to be contrasted to the effect of non-magnetic disorder in inducing the topological phase transitions in WSM, where in the framework of Born approximation, the renormalized topological mass is always made to be decreasing \cite{M.J.Park,H.Shapourian}.

In the following, we set the effective magnetic field $h_e>0$ and the resulted magnetization ${\cal M}_z<0$. Then the topological mass $m_1$ will be renormalized to its increased value $m_1'$, which will be further enhanced by the Hubbard interaction.

\subsection{Phase diagrams}

In Fig. 4(a) and (b), we plot the interaction-induced phase diagrams of WSM in parametric space $(m_1,U)$ and $(a_t,U)$, respectively.  In comparison with the noninteracing phase diagram in Fig. 2, we see that the Hubbard interaction can induce dramatic changes of the phase boundaries.  This is because, besides the topological mass renormalization, the interaction also renormalizes the tilting factor $\gamma_z$ and the Fermi velocity $v_z$ in the $z-$direction:
\begin{eqnarray}
&&\gamma_z'=a_t\Big|\frac{m_1-\frac{U}{2}{\cal M}_z}{t}\Big|,
\\
&&|v_z'|=t\sqrt{1-\Big(\frac{m_1-\frac{U}{2}{\cal M}_z}{t}\Big)^2}.
\end{eqnarray}

There are several features in the phase diagram in Fig. 4 that are worth pointing out.  First, when $m_1$ increases in Fig. 4(a), the phase boundaries deviate to the lower $U$ as ${\cal M}_z$ decreases with $m_1$, and when $a_t$ increases in Fig. 4(b), the phase boundaries deviate to higher $U$ as ${\cal M}_z$ increases with $a_t$.  Second, the phases of NI, WSM1 and WSM2 are all unstable to interaction.  If the interaction is strong enough, they will eventually be driven into the QAHI phase, in accordance with the previous analysis.  Third, when $m_1<t$, the system can be driven into WSM2 by Hubbard interaction as long as the tilting is nonvanishing, $a_t\neq0$.

For example, as along arrow $a$ in Fig. 4(a), when $|m_1'|<t$, the system initially lies in the NI phase.  Upon increasing $U$, the Weyl nodes move on $z-$axis, and correspondingly their tilting factor $\gamma_z'$ and Fermi velocity $v_z'$ change.  At $U=0.53$, the energy gap closes and the system enters WSM2 phase as the Weyl cones are overtilted as $\gamma_z'>|v_z'|$.  At $U=1.23$, $\gamma_z'$ begins to be smaller than $|v_z'|$, the topological phase transition from WSM2 to WSM1 happens.  When $\gamma_z'>|v_z'|$ at $U=4.04$, the Weyl cones are overtilted and the system enters the WSM2 phase again.  Finally, at $U=4.6$, the Weyl cones will meet at $k_z=0$ and annihilate.  As a result, the energy gap will be opened again, and the system is driven to the QAHI phase.

We also investigate the effect of thermal fluctuations caused by finite temperature on the phase diagram of WSM.  In Fig. 5, for cut at $m_1=-1.2$ in Fig. 4(a), along arrow $a$, we plot the interaction-induced phase diagram of WSM at finite temperature in the parametric space of $(U,$ ln$\beta)$ with $\beta=\frac{1}{k_BT}$.  It shows when the temperature is high (ln$\beta<0$), the thermal fluctuations will induce the larger critical interaction strength to drive both the metal-insulator and WSM1-WSM2 phase transitions.  With the further increase of temperature, the critical interactions tend to diverge.  When the temperature is low as ln$\beta>1$, the critical interaction for the transitions almost keep unchanged.  To make comparison to the archetypical Weyl material TaAs \cite{X.Huang}, we take $t=0.2$eV and the lattice constant $a=5\AA$, which lead to the Fermi velocity of $v=1$eV$\AA$ \cite{M.Udagawa}.  It can be estimated that ln$\beta=1$ corresponds to the real temperature of about 854K, which is much larger than the room temperature.  Therefore in 3D TaAs, the thermal fluctuations will be effectively frozen and has negligible effect on the Hubbard interaction-induced topological phased transitions.

\begin{figure}
\includegraphics[width=9.2cm]{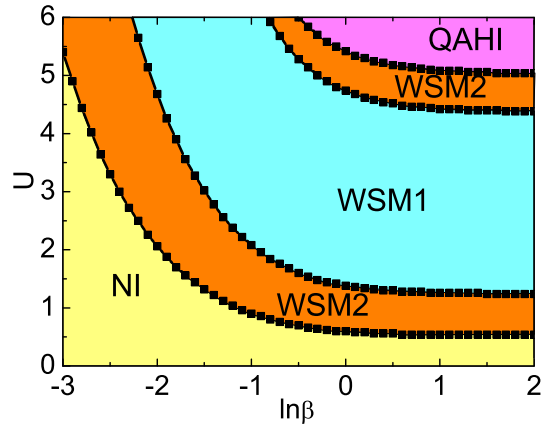}
\caption{(Color online) Interaction-induced phase diagrams of WSM at finite temperature, which is shown in the parametric space of $(U,$ ln$\beta)$ with $\beta=\frac{1}{k_BT}$.  The different phases are shown in different colors.}
\end{figure}

\subsection{Antiferromagnetic order}

Here we need to consider the magnetic property of the system, which is induced by the combined effects of the effective magnetic field $h_e$ and the Hubbard interaction.  On one hand, the same magnitude of $h_e$ on all sites prefers the FM order as to minimize the energy of the system.  On the other hand, when the system is at half-filling, the Hubbard interaction tends to induce the antiferromagnetic (AFM) order.  Therefore the two factors will compete with each other to determine the ground state of the system.

To find the AFM order, the unit cell that includes only one atom site needs to be enlarged to include more than one atom sites \cite{S.D.Matteo}.  As schematically shown in Fig. 6(a) of the lattice structure, the enlarged unit cell includes four atom sites of $A_1$, $B_1$, $A_2$ and $B_2$, with the ansatz of AFM-xyz order in all three directions.  Such a unit cell encloses four atomic sites that can be in principle inequivalent.  The enlarged unit cell in a cubic lattice structure may also be chosen in the $x-y$ plane or the $z-$direction (see Appendix) and the corresponding AFM orders are termed as AFM-xy or AFM-z.  Within the mean-field theory, we can calculate the magnetization on each site self-consistently and then judge the phase transition.

In the basis of $(c_{{\bf k}A_1\uparrow},c_{{\bf k}A_1\downarrow},
c_{{\bf k}B_1\uparrow},c_{{\bf k}B_1\downarrow},
c_{{\bf k}B_2\uparrow},c_{{\bf k}B_2\downarrow}$,
$c_{{\bf k}A_2\uparrow},c_{{\bf k}A_2\downarrow})$,
the $8\times8$ Hamiltonian describing the enlarged unit cell of the system becomes: \begin{widetext}
\begin{eqnarray}
&&H({\bf k})=
\nonumber\\
&&\begin{pmatrix}
\\
h_e& 0& -e^{ik_x}f_{\bf k}& e^{ik_x}g_{\bf k}&
e^{-ik_z}(p_{\bf k}+q_{\bf k})& 0& 0& 0
\\
0& -h_e& e^{ik_x}g^*_{\bf k}& e^{ik_x}f_{\bf k}&
0& e^{-ik_z}(-p_{\bf k}+q_{\bf k})& 0& 0
\\
-e^{-ik_x}f_{\bf k}& e^{-ik_x}g_{\bf k}& h_e& 0&
0& 0& e^{-ik_z}(p_{\bf k}+q_{\bf k})& 0
\\
e^{-ik_x}g^*_{\bf k}& e^{-ik_x}f_{\bf k}& 0& -h_e&
0& 0& 0& e^{-ik_z}(-p_{\bf k}+q_{\bf k})
\\
e^{ik_z}(p_{\bf k}+q_{\bf k})& 0& 0& 0&
h_e& 0& -e^{ik_x}f_{\bf k}& e^{ik_x}g_{\bf k}
\\
0& e^{ik_z}(-p_{\bf k}+q_{\bf k})& 0& 0&
0& -h_e& e^{ik_x}g^*_{\bf k}& e^{ik_x}f_{\bf k}
\\
0& 0& e^{ik_z}(p_{\bf k}+q_{\bf k})& 0&
-e^{-ik_x}f_{\bf k}& e^{-ik_x}g_{\bf k}& h_e& 0
\\
0& 0& 0& e^{ik_z}(-p_{\bf k}+q_{\bf k})&
e^{-ik_x}g^*_{\bf k}& e^{-ik_x}f_{\bf k}& 0& -h_e
\end{pmatrix},
\nonumber\\
\end{eqnarray}
\end{widetext}
with the parameters taken as $h_e=m_1+2m_0, f_{\bf k}=m_0(\text{cos}k_x+\text{cos}k_y),g_{\bf k}=t(\text{sin}k_x-i\text{sin}k_y),p_{\bf k}=t\text{cos}k_z, q_{\bf k}=a_t\text{sin}k_z$.

\begin{figure}
\includegraphics[width=8.6cm]{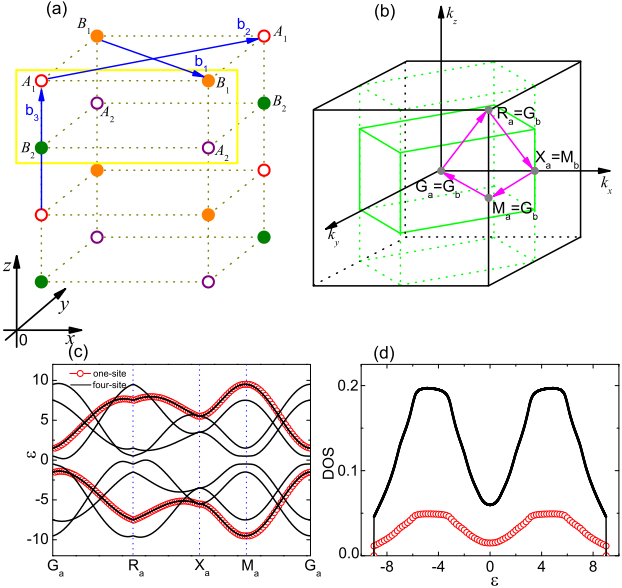}
\caption{(Color online) (a) Schematic plot of the lattice structure in 3D space. The yellow cell represents nearest-neighbor four-site cells in the $({\bf b}_1,{\bf b}_2,{\bf b}_3)$ basis.  (b) The first BZ for the one-(four-)site cell in the black (blue) lines.  (c) and (d) are the band structures and DOS per unit cell obtained from the tight-binding model, where (c) is calculated along the high-symmetric lines in the BZ, as shown by the purple lines in (b).  The other parameters are set as $a_t=1$, $m_0=2$ and $m_1=0.5$.}
\end{figure}

In Fig. 6(c), we plot the band structures along the high-symmetric lines $G_a-R_a-X_a-M_a-G_a$ in the 3D BZ (see Fig. 6(b)).  It shows that changing the choice of the unit cell will not change the band structures, but lead to the appearance of the additional bands.  In fact, the additional bands of the four-site cell are just the folding of the bands of the one-site cell, as $R_a=(\pi,\pi,\pi)$, $M_a=(\pi,\pi,0)$ in the BZ of the one-site cell both becomes equivalent to $\Gamma_b=(0,0,0)$ in the BZ of the four-site cell, and $X_a=(\pi,0,0)$ is equivalent to $M_b=(\pi,0,0)$.  The density of states (DOS) per unit cell is given as:
\begin{eqnarray}
\rho(\varepsilon)=\frac{1}{N}\sum_{\alpha,{\bf k}}
\delta[\varepsilon-\varepsilon_\alpha({\bf k})],
\end{eqnarray}
with $N$ being the number of unit cell and $\varepsilon_\alpha({\bf k})$ is the eigenenergy of $H({\bf k})$. In Fig. 6(d), the normalized DOS are shown, where the DOS of four-site cell are four times as those of one-site cell, as there are four atom sites in the enlarged unit cell.  The above analysis demonstrates that the enlarged unit cell constructed here is quite reliable and can be used for further calculations.

In Fig. 7, when the Hubbard interaction is strong, we plot the interaction-induced magnetic phase diagram with the same parameters as Fig. 4.  It shows that due to the competitions between $h_e$ and Hubbard $U$, when $U$ is below the critical interaction $U_c$, the FM order dominates as the magnetization ${\cal M}_{A_1z}={\cal M}_{A_2z}={\cal M}_{B_1z}={\cal M}_{B_2z}\sim-1$ and the system lies in the QAHI phase.  While when $U>U_c$, the AFM order dominates as ${\cal M}_{A_1z}={\cal M}_{A_2z}=-{\cal M}_{B_1z}=-{\cal M}_{B_2z}\sim=1$.  We can see that in Fig. 7(a), the phase boundary increases with $m_1$ while in Fig. 7(b), the phase boundary is less affected by the tilting parameter $a_t$. In both Fig. 7(a) and (b), a direct transition from FM order to AFM-xyz order can be seen, with the ordering vector of the spin density wave (SDW) as ${\bf Q}_1=(\pi,\pi,\pi)$.  It should be noted that the red solid (blue dotted) lines in both figures denote the separations between the QAHI and AFM-xy (AFM-z) phases.  However, when comparing the total energies of the ground states, the AFM-xyz order owns lower energy than the other two orders, so the AFM-xyz order is more energetically favorable and easily to be formed in this 3D WSM system.

\begin{figure}
\includegraphics[width=9.2cm]{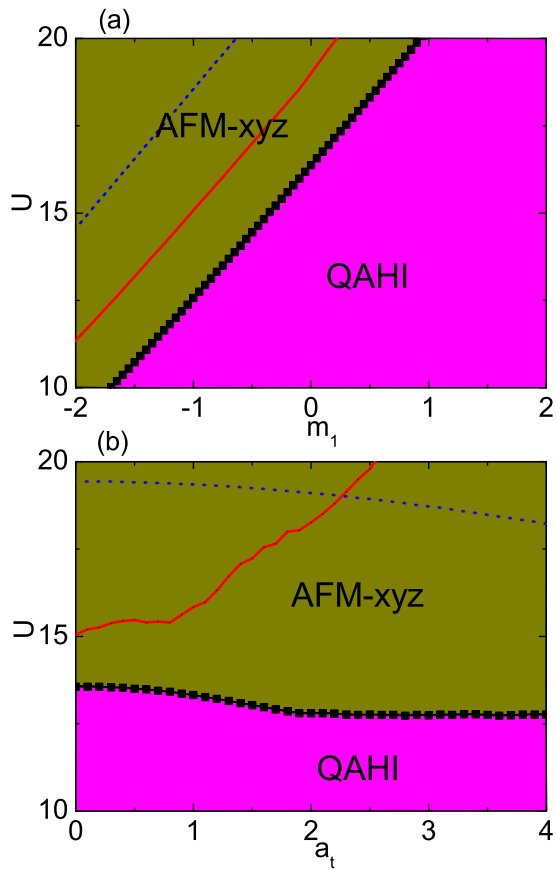}
\caption{(Color online) Interaction-induced phase diagrams in WSM at large$-U$ limit.  The red solid (blue dotted) lines in both figures are the separations between the QAHI and AFM-xy (AFM-z) phase, but actually these phases can not exist in the system, due to their higher ground state energies compared with AFM-xyz order.  The parameters are the same as Fig. 4.}
\end{figure}

The appearance of AFM order is supported by the previous studies of WSMs \cite{M.Laubach,J.Maciejko}, where the AFM order also exists when $U$ is strong.  The conclusion of SDW is in accordance with Ref. ~\cite{M.Laubach} using the variational cluster approach, but is different from Ref.~\cite{J.Maciejko} with the renormalization group analysis, where the ordering vector of the SDW is predicted to be equal to the momentum-space separation of the Weyl points.  So further theoretical and experimental studies to verify the SDW order are needed.

\subsection{Model of two pairs of Weyl nodes}

In this section, we study another WSM model where the inversion symmetry is broken but the TRS is preserved \cite{L.Huang,T.M.McCormick}:
\begin{eqnarray}
H_0'&=&t(\text{cos}k_x\sigma_x+\text{sin}k_y\sigma_y)+(m_1+t\text{cos}k_z)\sigma_z
\nonumber\\
&&+m_0(2-\text{sin}^2k_x-\text{cos}k_y)\sigma_z-\mu\sigma_0,
\end{eqnarray}
with the time-reversal operator ${\mathcal T}={\mathcal K}$ and ${\mathcal K}$ being the complex conjugation operator.  The peculiarity of this model is that when $|m_1|<t$, it hosts two pairs of Weyl nodes located at $E=0$ and ${\bf K}=(\pm\frac{\pi}{2},0,\pm Q)$, where $Q=$arccos($-\frac{m_1}{t}$).  Since the TRS is preserved, the Hall conductance vanishes.  So when $|m_1|>t$, the system does not have the QAHI phase, but lies in the NI phase, which is different from the model in Eq. (1).

When both the tilt and Hubbard interaction are included, the system becomes $H=H_0'+H_t+H_u$.  We focus on the small $U$ case.  The analysis and calculations show similar results for the MF parameters as in model (1).  Specifically, we find ${\cal M}_-=0$ and ${\cal M}_z<0$ for the positive effective magnetic field.  For ${\cal M}_-$ in Eq. (14), we have, for this model, $B_{\bf k}=t(\text{cos}k_x-i\text{sin}k_y)$ with the property of $B(k_x+\pi,k_y+\pi)=-B(k_x,k_y)$.  Therefore ${\cal M}_-$ vanishes as well when summing the momentum $\bf k$ over the BZ.  Qualitatively, however, numerical calculations (not shown) tell us that the same magnitude of interaction $U$ leads to smaller magnetization ${\cal M}_z$ in this model,  leading to the phase boundaries deviating to the larger Hubbard interactions.  This can be attributed to the enhancement of the itinerancy of electrons in the model of two-pair Weyl nodes, where, besides the nearest-neighbor hoppings, the next-nearest-neighbor hoppings also occur in the $x-$direction.  Therefore the density difference between two spin is weaken.  To compensate this, a larger Hubbard interaction is needed to induce the topological phase transitions.  In this sense, we suggest that the Hubbard interaction-induced mass renormalization and WSM1-WSM2 topological phase transitions have certain universality for the titled Weyl fermion systems.

\section{\label{sec:level1} Discussions and Summaries}

In conclusion, we have studied the effect of onsite Hubbard interaction on the phase diagrams of WSM with a nonzero tilt.  Within the MF theory, we self-consistently solve the MF parameters from the minimum model and then obtain the interaction-induced topological phase diagrams.  We find that the resultant renormalized topological mass can effectively change the Fermi velocity and the tilting of the Weyl cone.  As a result, the phase boundaries of both the metal-insulator phase transitions and WSM1-WSM2 phase transitions are renormalized.  We have checked that when the tilting term takes as higher order harmonics \cite{Y.Xu,M.Koshino}, similar results can also be obtained.  We have also analyzed the possible appearance of AFM orders at large$-U$ limit with the enlarged unit cell.  We hope the results can be validated in the known WSM materials \cite{L.Huang,K.Deng,J.Jiang,N.Xu,I.Beloposki,S.Y.Xu}, where the different phases can be characterized by their transport signatures, and in the cold-atom optical lattice experiment as well \cite{Y.Xu}.

We believe that the results of MF theory are qualitatively correct as the proper variations of the parameters with the interaction can be captured \cite{T.I.Vanhala}.  Thus the MF theory provides an intuitive understanding of the competition between the interaction and topology and can serve as a starting point for future studies.  To go beyond this and study the quantum fluctuations around the critical points, it would be interesting to use techniques such as the renormalization group or the quantum Monte Carlo methods, to investigate such problems.

\section{\label{sec:level1} Acknowledgements}

We would like to thank Yongping Zhang for helpful discussions.  This work was supported by NSF of Jiangsu Province of China (Grant No. BK20140129) and the Fundamental Research Funds for the Central Universities (Grant No. JUSRP51716A).

\section{Appendix}

Here we examine the other possible AFM orders due to the Hubbard interaction.

\subsection{Enlarged unit cell}

\begin{figure}
\includegraphics[width=9cm]{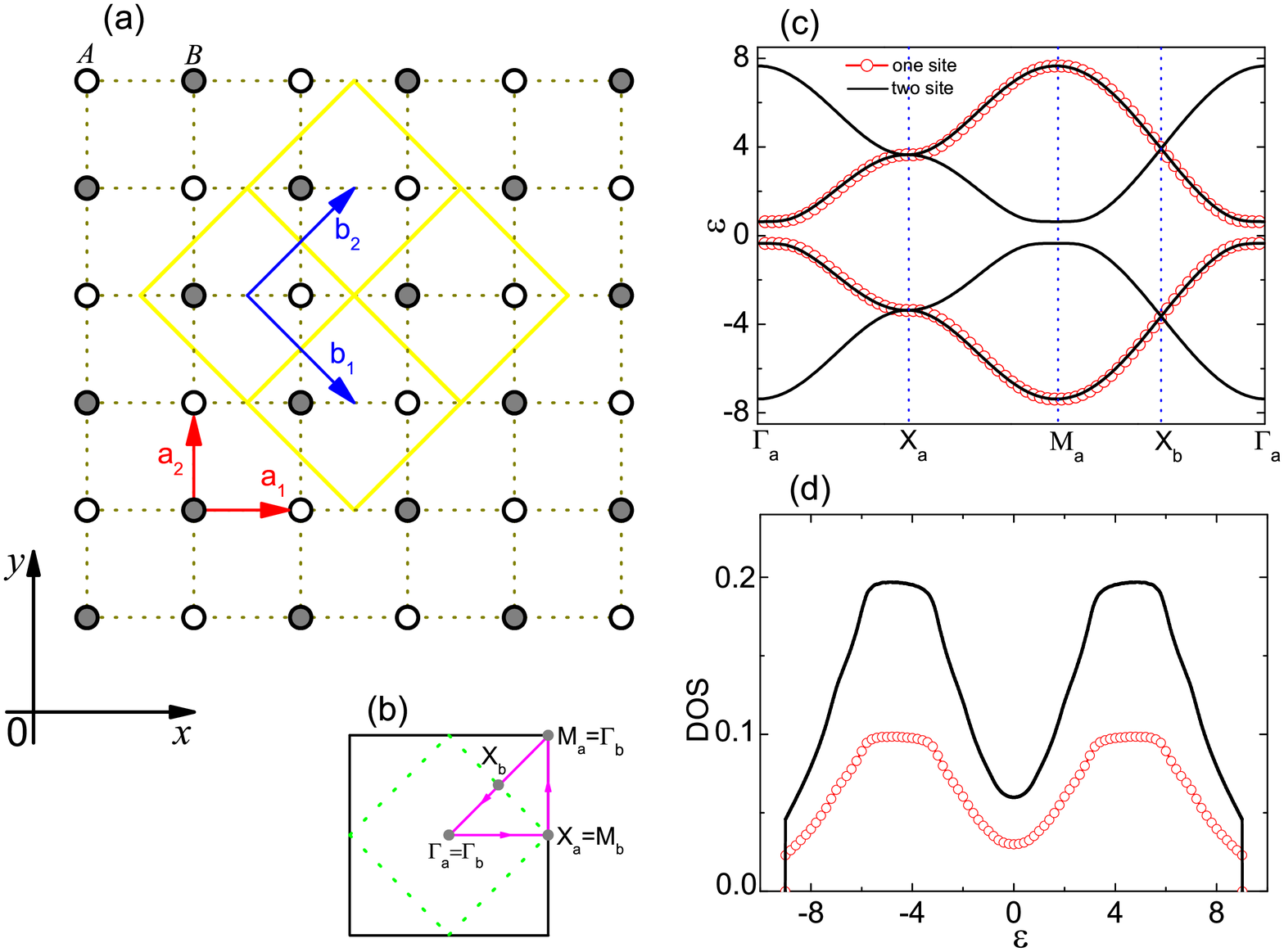}
\caption{(Color online) (a) Schematic plot of the lattice structure in the $x-y$ plane including one-site and two-site cells.  The four yellow cells represent nearest-neighbor cells in the $({\bf b}_1,{\bf b}_2)$ basis.  (b) The first $k_x-k_y$ BZ is plotted for the one-(two-)site cell in the solid black (dotted green) square.  (c) and (d) are the band structures and DOS per unit cell obtained from the tight-binding model, where (c) is calculated with $k_z=3$ and along the high-symmetric lines in the BZ, as shown by the purple lines in (b).  The other parameters are set as $a_t=1$, $m_0=2$ and $m_1=0.5$.}
\end{figure}

As the AFM order means the opposite spin orientations between neighboring sites, to find it, we need to enlarge the unit cell in the cubic lattice structure.  Besides the enlarged unit cell chosen in all three directions discussed in the main text, we can also choose the enlarged unit cell in the $x-y$ plane or in the $z-$direction and the corresponding AFM orders are termed as AFM-xy or AFM-z.

First, we consider the enlarged unit cell in the $x-y$ plane.  As shown in Fig. 8(a) of the lattice structure in the $x-y$ plane, the primitive one-site cell is spanned by the two vectors ${\bf a}_{1,2}$, while the enlarged two-site unit cell is spanned by two vectors ${\bf b}_{1,2}$ and are of double area.  In the basis of $(c_{{\bf k}A\uparrow},c_{{\bf k}A\downarrow},c_{{\bf k}B\uparrow}, c_{{\bf k}B\downarrow})^T$, the $4\times4$ Hamiltonian becomes:
\begin{eqnarray}
H({\bf k})=\begin{pmatrix}
d_{\bf k}+q_{\bf k}& 0& -e^{ik_x}f_{\bf k}& e^{ik_x}g_{\bf k}
\\
0& -d_{\bf k}+q_{\bf k}& e^{ik_x}g_{\bf k}^*& e^{ik_x}f_{\bf k}
\\
-e^{-ik_x}f_{\bf k}& e^{-ik_x}g_{\bf k}& d_{\bf k}+q_{\bf k}& 0
\\
e^{-ik_x}g_{\bf k}^*& e^{-ik_x}f_{\bf k}& 0& -d_{\bf k}+q_{\bf k}
\end{pmatrix},
\end{eqnarray}
with the parameters being the same as Eq. (19) and $d_{\bf k}=h_e+p_{\bf k}$.  After diagonalizing $H({\bf k})$, the energies are obtained as $\varepsilon_{\alpha\pm}({\bf k})=q_{\bf k}\pm\sqrt{|g_{\bf k}|^2+(d_{\bf k}+\alpha f_{\bf k})^2}$, with $\alpha=\pm1$.

In Fig. 8(c), we plot the band structures along the high-symmetric lines $\Gamma_a-X_a-M_a-X_b-\Gamma_a$ in the $k_x-k_y$ BZ (see Fig. 8(b)).  It shows that changing the choice of the unit cell will not change the band structures, as the additional bands of the two-site cell are just the folding of the bands of the one-site cell.  This is because the point $M_a=(\pi,\pi)$ in the BZ of the one-site cell becomes equivalent to $\Gamma_b=(0,0)$ in the BZ of the two-site cell.  In Fig. 8(d), the normalized DOS are shown, where the DOS of two-site cell are double those of one-site cell, as there are two atom sites per unit cell in the case of two-site cell.

\begin{figure}
\includegraphics[width=9cm]{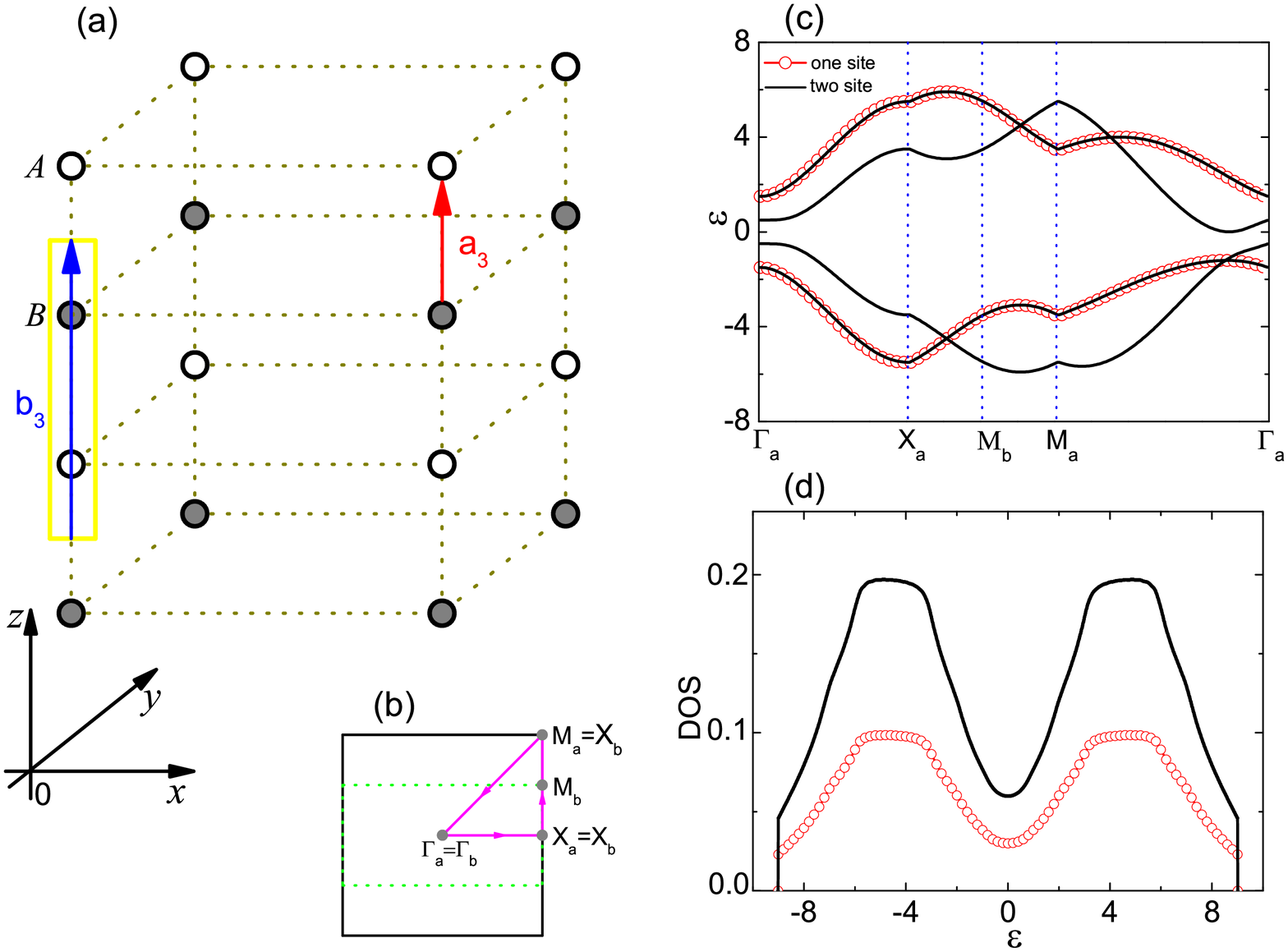}
\caption{(Color online) (a) Schematic plot of the lattice structure in the 3D space, including one-site and two-site cells in $z-$direction.  The yellow cells represent nearest-neighbor cells.  (b) The first $k_y-k_z$ BZ is plotted for the one-(two-)site unit cell in the solid black (dotted green) square.  (c) and (d) are the band structures and DOS per unit cell obtained from the tight-binding model, where (c) is calculated when $k_x=0$ and along the high-symmetric lines in the BZ, as shown by the purple lines in (b).  The other parameters are the same as Fig. 7. }
\end{figure}

\begin{figure}
	\includegraphics[width=8.8cm]{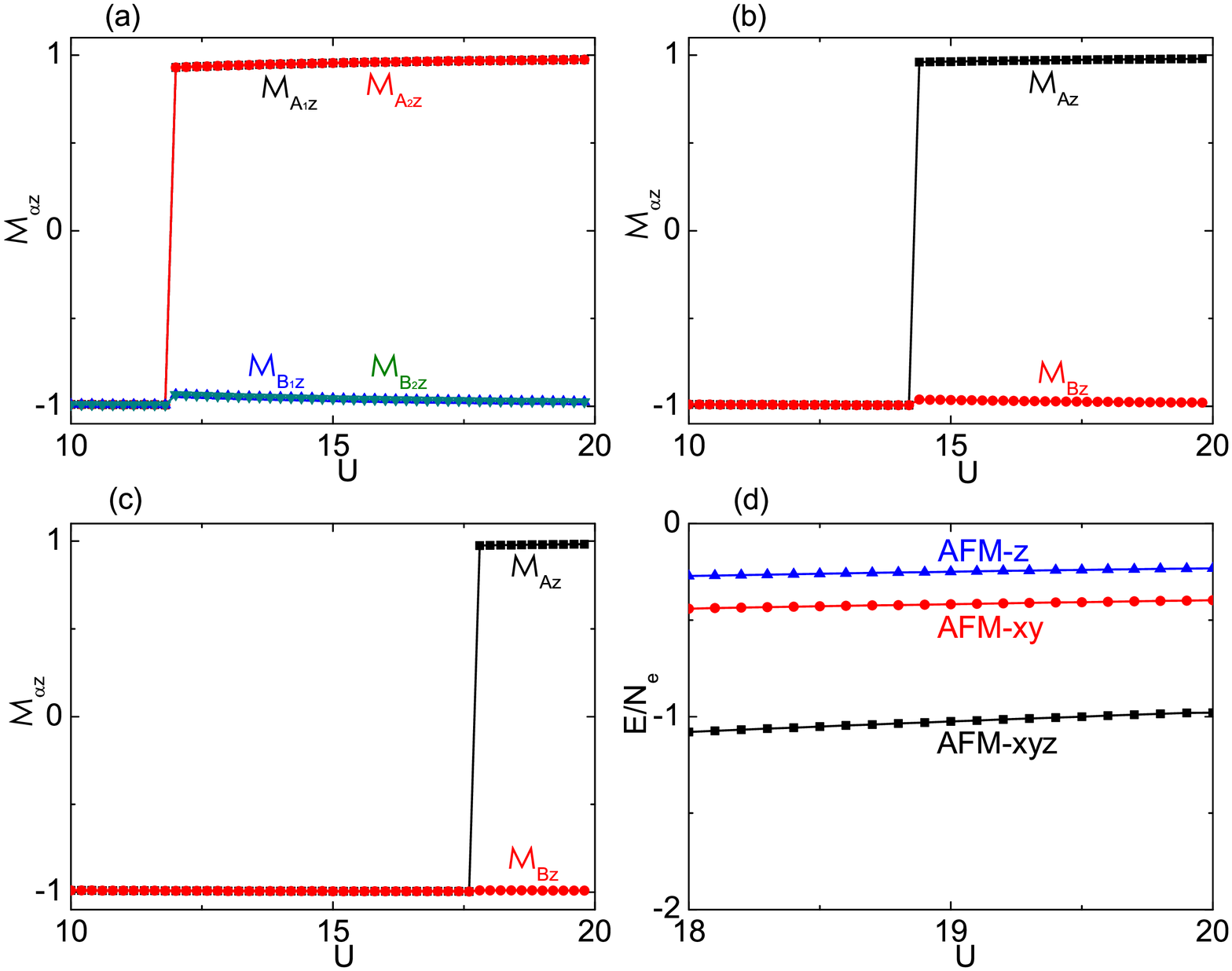}
	\caption{(Color online) Plot of the magnetization ${\cal M}_{\alpha z}$ vs the Hubbard interaction $U$ for the enlarged unit cell in the $x-y$ plane (a) and $z-$direction (b) and all three directions (c).  The parameters are chosen as $m_0=2$, $m_1=-1.2$ and $a_t=1$.  (d) shows the total energy of the ground state $E/N_e$ for different AFM orders. }
\end{figure}

We can also choose the enlarged unit cell in the $z-$direction, as plotted in Fig. 9(a) of the lattice structure, where the vectors ${\bf a}_3$ of one-site cell and ${\bf b}_3$ of two-site cell are shown.  In $\bf k-$space, the $4\times4$ Hamiltonian is written as:
\begin{eqnarray}
&&H({\bf k})=
\nonumber\\
&&\begin{pmatrix}
h_e-f_{\bf k}& g_{\bf k}& e^{-ik_z}(p_{\bf k}+q_{\bf k})& 0
\\
g^*_{\bf k}& -(h_e-f_{\bf k})& 0&  e^{-ik_z}(-p_{\bf k}+q_{\bf k})
\\
e^{ik_z}(p_{\bf k}+q_{\bf k})& 0& h_e-f_{\bf k}& g_{\bf k}
\\
0& e^{ik_z}(-p_{\bf k}+q_{\bf k})& g_{\bf k}^*& -(h_e-f_{\bf k})
\end{pmatrix},
\nonumber\\
\end{eqnarray}
with the parameters being the same as Eq. (19).  The energies can be solved directly as $\varepsilon_{\alpha\pm}({\bf k})=\alpha q_{\bf k}\pm\sqrt{(h_e-f_{\bf k}+\alpha p_{\bf k})^2+|g_{\bf k}|^2}$, with $\alpha=\pm$.

In Fig. 9(c), we plot the band structures along the high-symmetric lines $\Gamma_b-X_b-M_b-X_b-\Gamma_b$ in the $k_y-k_z$ BZ (see Fig. 9(b)).  The enlarged unit cell do not change the band structures as well and the additional bands are just the folding of the bands of the one-site cell.  It should be noted that the folded BZ is different from previous one as the enlarged unit cell are chosen in different directions.  In this case, the point of $M_a=(\pi,\pi)$ in the BZ of the one-site cell becomes equivalent to $X_b=(\pi,0)$ in the BZ of two-site cell.  In Fig. 9(d), the DOS per unit cell is plotted, where the enlarged unit cell also has the twice DOS of the one-site cell.

These analysis demonstrate that the enlarged unit cells constructed by different choices are quite reliable and reasonable.

\subsection{Mean-field theory}

Within the mean-field approximation, the Hubbard interaction for the enlarged unit cell in the momentum space is given as:
\begin{eqnarray}
H_U=U\sum_{{\bf k},\alpha}\Big[\langle n_{\alpha\downarrow}\rangle n_{{\bf k}\alpha\uparrow}
+\langle n_{\alpha\uparrow}\rangle n_{{\bf k}\alpha\downarrow}
-\langle n_{\alpha\uparrow}\rangle\langle n_{\alpha\downarrow}\rangle\Big],
\end{eqnarray}
in which the index $\alpha=A,B$ for AFM-xy and AFM-z and $\alpha=A_1,B_1,A_2,B_2$ for AFM-xyz.  Here we have kept the constant term, which does not depend on the creation or annihilation operators but only on their average values.  This term must be included in calculating the total energy of the system as to help judge the ground state.  We can define the mean-field parameters of the charge density and magnetization on site $\alpha$ as \cite{V.S.Arun}:
$\rho_\alpha=\langle n_{\alpha\uparrow}\rangle+\langle n_{\alpha\uparrow}\rangle$,
${\cal M}_{\alpha z}=\langle n_{\alpha\uparrow}\rangle-\langle n_{\alpha\downarrow}\rangle$.
When the system is at half-filling as we have chosen before, the charge densities on each atom site are naturally $\rho_\alpha=1$.  ${\cal M}_{\alpha z}$ on each site $\alpha$ can be calculated by the self-consistent iterative algorithm.  We have carefully checked the results for different size of the cubic system as $L=10,20,30$, which exhibit good convergence.

In Fig. 10(a)-(c), as along the arrow in Fig. 4(a), we plot the magnetization ${\cal M}_{\alpha z}$ vs the Hubbard interaction $U$ for different AFM orders.  It can be clearly seen that as the interaction is strong enough, the magnetization on each site can reach its saturation value of $-1$.  In Fig. 10(a), when the Hubbard interaction is below the critical interaction $U<U_c=11.82$, the FM order dominates as ${\cal M}_{\alpha z}\sim-1$ and when $U>U_c$, the AFM order appears as ${\cal M}_{A_1z}={\cal M}_{A_2z}=-{\cal M}_{B_1z}=-{\cal M}_{B_2z}\sim1$.  It shows that during the phase transition, the magnetization ${\cal M}_{B_1z}$ and ${\cal M}_{B_2z}$ keeps almost unchanged, while ${\cal M}_{A_1z}$ and ${\cal M}_{A_2z}$ show an abrupt change, pointing to a first-order phase transition.  While in Fig. 10(b) and (c), the critical $U_c$ for the appearance of AFM-xy and AFM-z are 14.31 and 17.75, respectively.  In Fig. 10(d), we plot the total energy of different AFM ground states, where we have used $E/N_e$ with $N_e$ being the electron number instead of $E$ as to avoid the effect of the unit cell size.  It shows clearly that the AFM-xyz order owns the lower energy than the other two orders and therefore is more energetically favorable.

\end{document}